\documentclass[12pt]{article}

\newcommand{\startappendix}{\appendix\setcounter{equation}{0}
  \renewcommand{\theequation}{\Alph{section}.\arabic{equation}}}
\newcommand{\beq}{\begin{eqnarray}}
\newcommand{\eeq}{\end{eqnarray}}
\newcommand{\beqnn}{\begin{eqnarray*}}
\newcommand{\eeqnn}{\end{eqnarray*}}
\newtheorem{theorem}{Theorem}

\newtheorem{lemma}{Lemma}

\newcommand{\proof}{\medskip\par\noindent {\it Proof.\/}\quad}
\newcommand{\qed}{{\it Q.E.D.\/} \bigskip\par}

\newcommand{\rd}{\partial}

\newcommand{\Tr}{\mathop{\mathrm{Tr}}}
\newcommand{\diag}{\mathop{\mathrm{diag}}}
\newcommand{\rmSU}{\mathrm{SU}}
\newcommand{\rmsu}{\mathrm{su}}
\newcommand{\otcomma}{\stackrel{\otimes}{,}}
\newcommand{\Res}{\mathop{\mathrm{Res}}}

\newcommand{\ZZ}{\mathbf{Z}}
\newcommand{\calH}{\mathcal{H}}
\newcommand{\calO}{\mathcal{O}}
\begin{document}

%%%%%%%%%%%%%%%%
%% title page %% 
%%%%%%%%%%%%%%%%

\title{Spectral curve and Hamiltonian structure 
of isomonodromic SU(2) Calogero-Gaudin system}
\author{Kanehisa Takasaki\\
{\normalsize Department of Fundamental Sciences}\\
{\normalsize Faculty of Integrated Human Studies, Kyoto University}\\
{\normalsize Yoshida, Sakyo-ku, Kyoto 606-8501, Japan}\\
{\normalsize E-mail: takasaki@math.h.kyoto-u.ac.jp}}
\date{}
\maketitle

\begin{abstract}
This paper presents a new approach to the Hamiltonian 
structure of isomonodromic deformations of a matrix system 
of ODE's on a torus.  An isomonodromic analogue of the 
$\rmSU(2)$ Calogero-Gaudin system is used for a case study 
of this approach.  A clue of this approach is a mapping 
to a finite number of points on the spectral curve of 
the isomonodromic Lax equation.  The coordinates of 
these moving points give a new set of Darboux coordinates 
called the spectral Darboux coordinates.  The system of 
isomonodromic deformations is thereby converted to a 
non-autonomous Hamiltonian system in the spectral Darboux 
coordinates.  The Hamiltonians turn out to resemble those 
of a previously known isomonodromic system of a second 
order scalar ODE.  The two isomonodromic systems are shown 
to be linked by a simple relation.  
\end{abstract}
\begin{flushleft}
arXiv:nlin.SI/0111019
\end{flushleft}
%\begin{flushleft}
%Running title: Spectral curve and Hamiltonian structure 
%\end{flushleft}

\newpage

%%%%%%%%%%%%%%%
%% comment page
%%%%%%%%%%%%%%%

\section*{Comments on revised version}

This is a revised version of the paper published in 
Journal of Mathematical Physics vol. 44, no. 2 (2003), 
pp. 3979--3999.  It turned out after publication 
that the published version contains serious errors 
on the construction of the Lax and zero-curvature equations.  
Namely, the $M$-matrices $M_j(z)$ have to be corrected 
by an extra diagonal matrix $\diag(p_j,-p_j)$; 
the zero-curvature equations without these terms lead to 
a contradiction. The emergence of these correction terms 
was already pointed out in the paper of Korotkin and Samtleben 
\cite{bib:Ko-Sa}.  As they stressed therein, this problem 
stems from the fact that the isomonodromic system in question 
is a constrained system.  The quantities $p_j$, $j = 1,\ldots,N$, 
may be interpreted as the Lagrange multipliers in 
the Hamiltonian formulation of this constrained system.  
This issue is also partly related to the treatment of 
the coefficient $\kappa$ that arises in the construction 
of spectral Darboux coordinates.  

To correct these errors, Section III has been fully revised. 
Firstly, the ``fake'' Hamiltonians $\calH_j$ are replaced 
by the Hamiltonians $\widetilde{\calH}_j$ with a correction 
term proportional to the left hand side of the constraint.  
The equations of motion of the Calogero variables $q,p$ and 
the spin variables $A_k$ are defined by these Hamiltonians.  
The $t$-dependence of $p_j$'s is not determined at this stage.  
Secondly, the $M$-matrices $M_j(z)$ are corrected by the diagonal matrix 
$\diag(p_j,-p_j)$.  The Lax and zero-curvature equations are 
reformulated in terms of the corrected $M$-matrices $\widetilde{M}_j(z)$.  
The diagonal part of the zero-curvature equations then yields 
a set of new differential equations for $p_j$'s.  As it turns out, 
the equations of motion of $q,p$ and $A_k$'s ensure the integrability, 
in the sense of Frobenius, of these equations.  Thus, as Korotkin and 
Samtleben \cite{bib:Ko-Sa} noted, the $t$-dependence of $p_j$'s 
are eventually determined by the zero-curvature equations. 

The subsequent sections are left mostly intact, except for 
the last part of Section VI that contained a wrong statement 
on $\kappa$.  

I would like to express my gratitude to Henning Samtleben 
for helpful suggestions on this issue.

\newpage

%%%%%%%%%%%%%%%
%% main text %% 
%%%%%%%%%%%%%%%

\section{Introduction}

The notion of isomonodromic deformations originates 
in the celebrated work of R. Fuchs \cite{bib:Fu}. 
Fuchs studied isomonodromic deformations of a second 
order linear ODE of the form 
\beqnn
    \frac{d^2y}{dz^2} 
  + \left(\frac{a}{z^2} + \frac{b}{(z-1)^2} 
      + \frac{c}{(z-t)^2} + \frac{d}{z(z-1)} 
      - \frac{3}{4(z - \lambda)^2} 
      \right. \\
      \left. \mbox{} - \frac{t(t-1)K}{z(z-1)(z-t)} 
      + \frac{\lambda(\lambda - 1)\nu}{z(z-1)(z-\lambda)} 
    \right) y = 0 
\eeqnn
with five regular singular points $z = 0,1,\infty,t,\lambda$ 
on the Riemann sphere, and discovered a nonlinear ODE 
that is nowadays called the sixth Painlev\'e equation.  
His work was soon generalized by Garnier \cite{bib:Ga1} 
and Schlesinger \cite{bib:Sch} in two different directions.  
Whereas Garnier extended the work of Fuchs to a second order 
linear ODE with more singularities (including irregular ones 
as well), Schlesinger studied a matrix system of the form 
\beqnn
    \frac{dY}{dz} = \sum_{j=1}^N \frac{A_j}{z-t_j}Y, 
\eeqnn
and obtained the so called Schlesinger system 
\beqnn
    \frac{\rd A_k}{\rd t_j} 
    = (1 - \delta_{jk}) \frac{[A_k,A_j]}{t_k-t_j} 
    - \delta_{jk}\sum_{l\not= k}\frac{[A_k,A_l]}{t_k-t_l} 
\eeqnn
that characterizes isomonodromic deformations.  
It turned out afterwards \cite{bib:Ga2} that Garnier's 
isomonodromic deformations with an arbitrary number 
of regular singular points can be reconstructed from 
the $2 \times 2$ Schlesinger system. 

The next stage of generalization is, naturally, 
isomonodromic deformations on a torus.  This issue 
was first tackled by Okamoto \cite{bib:Ok-ell1}, 
who obtained a system of isomonodromic deformations 
of a second order scalar ODE on a torus.  One of 
his remarkable results is that the isomonodromic 
system on a torus can be formulated as a Hamiltonian 
system in much the same way as Garnier's isomonodromic 
system on a sphere was converted to a Hamiltonian 
system \cite{bib:Ok-gar}. Iwasaki \cite{bib:Iw} extended 
Okamoto's work to scalar ODE's of an arbitrary order 
on an arbitrary compact Riemann surface, and elucidated 
the geometric origin of the Hamiltonian structure that 
Okamoto derived.  The study of isomonodromic systems 
on a torus was further refined by Okamoto himself 
\cite{bib:Ok-ell2} and Kawai \cite{bib:Ka}. 

As regards matrix systems, Korotkin and Samtleben 
\cite{bib:Ko-Sa} constructed an example of isomonodromic 
deformations of a $2 \times 2$ matrix system on a torus. 
Levin and Olshanetsky \cite{bib:Le-Ol} developed a general 
framework in which the Schlesinger system and Korotkin 
and Samtleben's isomonodromic system are placed, along 
with generalizations to higher genus Riemann surfaces, 
in a unified way.  Some more examples of matrix systems 
with different structures are also known 
\cite{bib:Ta-gau,bib:Ta-ecm,bib:Ko-Ma-Sa,bib:Kr-im}. 
Compared with Okamoto and Iwasaki's formulation, these 
``elliptic analogues of the Schlesinger system'' are 
obtained on an entirely different ground, such as conformal 
field theories, vector bundles on a torus, KZ equations, 
and (classical or quantum) integrable systems. 
This can be seen in the structure of the matrix 
linear system 
\beqnn
    \frac{dY}{dz} = L(z)Y 
\eeqnn
for which isomonodromic deformations are constructed. 
Namely, the matrix $L(z)$ (``$L$-matrix'') in these 
examples is borrowed from the isospectral Lax equation 
of an integrable system, though the Lax equation of 
isomonodromic deformations takes the non-isospectral form 
\beqnn
    \frac{\rd L(z)}{\rd t_j} 
  = [L(z), M_j(z)] - \frac{\rd M_j(z)}{\rd z}. 
\eeqnn
Each of those isomonodromic systems is thus accompanied 
by an isospectral partner.  

The correspondence between isospectral and isomonodromic 
systems will have a number of significant implications.  
Among them, we are particularly interested in the role 
of ``spectral Darboux coordinates''.  The notion of 
spectral Darboux coordinates was introduced by 
the Montreal group for isospectral systems with 
a rational $L$-matrix \cite{bib:AHH}, 
and extended to isomonodromic systems on a sphere 
\cite{bib:Ha-dual,bib:Ha-Wi}.  As they demonstrated 
for those cases, one can construct a mapping from 
the Lax equation of this type of systems to a dynamical 
system of a finite number of points $P_1,\ldots,P_N$ on 
the spectral curve 
\beqnn
    \Gamma = \{(z,w) \mid \det(wI - L(z)) = 0\}, 
\eeqnn
though the spectral curve itself becomes dynamical 
in the case of isomonodromic deformations.  
Spectral Darboux coordinates are the coordinates 
$\lambda_1,\ldots,\lambda_N, \mu_1,\ldots,\mu_N$ 
of the moving points $P_k = (\lambda_k,\mu_k)$.  
These coordinates lead to ``separation of variables'' 
of isospectral systems.  The most classical case is 
the so called Moser systems \cite{bib:Mo}; 
separation of variable of those systems was 
worked out by Moser himself.   Remarkably, 
as Harnad and Wisse pointed out \cite{bib:Ha-Wi}, 
almost the same story repeats on the isomonodromic side, 
except that separability is lost there.  In particular, 
this shows an algebro-geometric interpretation of 
Okamoto's reformulation \cite{bib:Ok-gar} of 
Garnier's work \cite{bib:Ga2} on the $2 \times 2$ 
Schlesinger system.  

This paper presents a similar approach to one of 
the ``elliptic analogues'' of the Schlesinger systems, 
namely, the aforementioned isomonodromic system of 
Korotkin and Samtleben (in a slightly modified form).  
The isospectral partner of this isomonodromic system 
is the Calogero-Gaudin system \cite{bib:Ne,bib:En-Ru} 
for the $\rmSU(2)$ group.  Separation of variables of 
the usual $\rmSU(2)$ Calogero-Gaudin system has been 
developed by Brzezi\'nski \cite{bib:Br} and Enriquez et al. 
\cite{bib:En-Fe-Ru} (including ``quantum separation of 
variables'' in the sense of Sklyanin \cite{bib:Sk-sov}). 
Our method is more or less parallel to theirs, in particular, 
that of Brzezi\'nski.  Actually, it is a rational 
(rather than elliptic) model of the $\rmSU(2)$ 
Calogero-Gaudin system that he considered.  
Thus we are to extend his method in two-fold ways --- 
firstly, to an elliptic model (which is the subject 
of the work of Enriquez et al. as well), and secondly, 
to an isomonodromic system.   

A main outcome of our consideration (summarized in Theorem 1, 
Section V) is that the isomonodromic $\rmSU(2)$ Calogero-Moser 
system can be converted to a non-autonomous Hamiltonian system 
in the spectral Darboux coordinates.  The Hamiltonians of 
this non-autonomous system turn out to be a considerably 
intricate functions of the Darboux coordinates and 
the time variables.  Remarkably, however, a very similar 
Hamiltonian system has been discovered by Okamoto 
\cite{bib:Ok-ell2} for isomonodromic deformations of 
a second order scalar ODE on a torus.  We shall show 
a natural explanation of this coincidence from our pouint 
of view. 

This paper is organized as follows.  Sections II and III 
are for preparation.  In Section II, the Poisson structure 
of the $L$-matrix of the Calogero-Gaudin systems is reviewed. 
In Section III, the isomonodromic system is formulated 
in terms of two canonically conjugate ``Calogero variables'' 
and a set of ``spin variables''.  Section IV and V 
are the main part of this paper.  In Section IV, 
the spectral curve and the spectral Darboux coordinates 
are introduced.  In Section V, the non-autonomous 
Hamiltonian system is derived.  Section VI deals 
with the relation to isomonodromic deformations of 
a second order scalar ODE.  Section VII is for conclusion 
and  supplementary remarks.  Part of technical details 
are collected in Appendices.

\section{$L$-matrix and Poisson structure}

\subsection{$L$-matrix}

Following the idea of Korotkin and Samtleben
\cite{bib:Ko-Sa}, we start from the $L$-matrix 
\beq
    L(z) 
    = \left(\begin{array}{cc} 
      p & 0 \\
      0 & -p 
      \end{array}\right) 
    + \sum_{j=1}^N 
      \left(\begin{array}{cc}
      \zeta(z-t_j)A_j^3 & \phi(q,z-t_j)A_j^{-} \\
      \phi(-q,z-t_j)A_j^{+} & - \zeta(z-t_j)A_j^3 
      \end{array}\right), 
\eeq
where $q$ and $p$ are Calogero variables, 
$A_j^{\pm}$ and $A_j^3$ are spin variables, 
$\zeta(z)$ denotes the Weierstrass $\zeta$ 
function and $\phi(u,z)$ the auxiliary 
function that is widely used in the study 
of systems of the Calogero type: 
\beq
    \zeta(z) = \frac{\sigma'(z)}{\sigma(z)}, \quad 
    \phi(u,z) = \frac{\sigma(u-z)}{\sigma(u)\sigma(z)}. 
\eeq
Here $\sigma(z)$ is Weierstrass sigma function, 
and the prime stands for a derivative, 
i.e., $\sigma'(z) = d\sigma(z)/dz$. 
Let $2\omega_1$ and $2\omega_3$ denote the primitive 
periods of the Weierstrass functions.  Throughout 
this paper, we assume that $t_j \not= t_k$ if 
$j\not= k$. 

This $L$-matrix is slightly different from that 
of Korotkin and Samtleben \cite{bib:Ko-Sa}.  
They use Jacobi's elliptic theta function 
$\vartheta_1$ rather than Weierstrass' sigma 
function $\sigma$. Their $L$-matrix is thereby 
more suited for formulating isomonodromic 
deformations against the modulus $\tau$. We dare 
to modify Korotkin and Samtleben's $L$-matrix 
because this simplifies the use of interpolation 
formulae of elliptic functions.  It should be possible 
to start from the $L$-matrix of Korotkin and Samtleben 
and to derive substantially the same results, though 
we shall not pursue it in this paper.  

The Poisson structure of the dynamical variables 
is a standard one.  The Calogero variables $q,p$ 
are, in fact, the relative coordinate $q_1 - q_2$ 
and momentum $(p_1 - p_2)/2$ of a two body system 
with canonical variables $(q_1,q_2,p_1,p_2)$, and 
become a canonically conjugate pair $\{q,p\} = 1$ 
in themselves.  The spin variables $A_j^\pm,A_j^3$ 
obey the $\rmsu(2)$ relations 
\beq
    \{A_j^3, A_k^{\pm}\} = \pm \delta_{jk} A_k^\pm, \quad 
    \{A_j^{+}, A_k^{-}\} = 2\delta_{jk}A_k^3 
\eeq
with respect to the Poisson bracket.  

The Poisson bracket of the spin variables is nothing 
but the Kostant-Killilov bracket for the residue matrix 
\beq
    A_j = \left(\begin{array}{cc}
          A_j^3 & A_j^{-} \\
          A_j^{+} & - A_j^3
          \end{array}\right)
\eeq
of $L(z)$ at $z = t_j$.  The conjugacy class
\beq
    \calO_j = \{A_j \mid  A_j \sim \diag(\theta_j/2,-\theta_j/2)\}
\eeq
of semi-simple matrices with fixed eigenvalues $\pm\theta_j/2$ 
is a maximal (two-dimensional) symplectic leaf of this 
Poisson structure.  One can use a canonically conjugate 
pair $(x_j,\xi_j)$, $\{x_j,\xi_j\} = 1$, to parametrize 
this symplectic leaf as follows: 
\beq
    A_j^{+} = - \frac{\xi_j^2}{2} + \frac{\theta_j^2}{2x_j^2}, 
    \quad 
    A_j^{-} = \frac{x_j^2}{2}, 
    \quad 
    A_j^3 = \frac{x_j\xi_j}{2}. 
    \label{eq:A-by-xxi}
\eeq
Note that this parametrization is consistent with the 
Poisson bracket of $A_j^{\pm,3}$.

\subsection{Poisson bracket of $L$-matrix elements}

Let us write the matrix elements of $L(z)$ as 
\beq
    L(z) = \left(\begin{array}{cc}
           A(u) & B(u) \\
           C(u) & -A(u)
           \end{array}\right). 
\eeq
More explicitly, 
\beqnn
    A(u) &=& p + \sum_{j=1}^N \zeta(z-t_j)A_j^3, 
    \\
    B(u) &=& \sum_{j=1}^N \phi(q,z-t_j)A_j^{-}, 
    \\
    C(u) &=& \sum_{j=1}^N \phi(-q,z-t_j)A_j^{+}. 
\eeqnn
The non-zero Poisson brackets of these matrix elements 
take the form
\beq
    \{A(z),B(w)\} &=& B(z)\phi(-q,z-w) - B(w)\zeta(z-w), 
    \\
    \{A(z),C(w)\} &=& - C(z)\phi(q,z-w) + C(w)\zeta(z-w), 
    \\
    \{B(z),C(w)\} &=& 2(A(z) - A(w))\phi(q,z-w) 
                    + 2\phi_u(q,z-w)\sum_{j=1}^N A_j^3. 
\eeq
where 
\beqnn
    \phi_u(u,z) 
    = \frac{\rd \phi(u,z)}{\rd u} 
    = - \phi(u,z)(\zeta(z - u) + \zeta(u)). 
\eeqnn
Thus the Poisson algebra of the matrix elements of $L(z)$ 
{\it almost closes} up to the extra term proportional 
to $\sum_{j=1}^N A_j^3$, which is later set to zero in order 
to derive the Lax equation.  

These Poisson commutation relations can be easily 
verified by direct calculations using the functional identity 
\beq
    \phi(u,z)\phi(-u,w) + \phi(u,z-w)(\zeta(z) - \zeta(w)) 
    + \phi_u(u,z-w) = 0 
    \label{eq:phi(u,z)phi(-u,z)}
\eeq
of the auxiliary functions.  This functional identity is 
a consequence of the more general one 
\beq
    \phi(u,z)\phi(v,w) + \phi(u+v,z)\phi(-v,z-w) 
    - \phi(u+v,w)\phi(u,z-w)= 0, 
    \label{eq:phi(u,z)phi(v,w)} 
\eeq
from which the former identity can be derived by 
letting $v \to -u$.  

The Poisson structure of the $L(z)$-matrix elements 
can be cast into the compact form 
\beq
    \{L(z) \otcomma L(w)\} 
    &=& \sum_{a,b,c,d}\{L_{ab}(z),L_{cd}(w)\}
        E_{ab} \otimes E_{cd}  
    \nonumber \\
    &=& [L(z) \otimes I +  I \otimes L(w), \ r(z-w)] 
    + 2 \frac{\rd r(z-w)}{\rd q}\sum_{j=1}^N A_j^3. 
    \label{eq:PCR-for-L}
\eeq
where $E_{ab}$ denotes the matrix with the $(a,b)$ 
element equal to $1$ and the other elements vanishing.  
The $r$-matrix takes the form 
\beq
    r(z-w) &=& \zeta(z-w)E_{11} \otimes E_{11} 
           + \phi(q,z-w)E_{12} \otimes E_{21} 
           \nonumber \\
           && \mbox{}
           + \phi(-q,z-w)E_{21} \otimes E_{12} 
           + \zeta(z-w)E_{22} \otimes E_{22} 
           \nonumber \\
           &=& 
             \left(\begin{array}{cccc}
             \zeta(z-w) & 0 & 0 & 0 \\
             0 & 0 & \phi(q,z-w) & 0 \\
             0 & \phi(-q,z-w) & 0 & 0 \\
             0 & 0 & 0 & \zeta(z-w)
             \end{array}\right), 
\eeq
which is a special case of the well known dynamical 
$r$-matrix of the elliptic Calogero-Moser system 
\cite{bib:Av-Ta,bib:Sk-ecm,bib:Br-Su}.

\section{Hamiltonians and Lax equations}

\subsection{Hamiltonians}

The fundamental Poisson commutation relation 
(\ref{eq:PCR-for-L}) implies that the standard 
quadratic Hamiltonians 
\beq
    \calH_j 
    = \Res_{z=t_j}\frac{1}{2}\Tr L(z)^2 
\eeq
are not Poisson-commutative in themselves, but 
commute up to a term proportional to 
$\sum_{j=1}^N A_j^3$: 
\beq
    \{\calH_j, \calH_k\} \propto \sum_{j=1}^N A_j^3. 
\eeq
The factor $\sum_{j=1}^N A_j^3$ itself commutes 
with $\calH_j$'s: 
\beq
    \Bigl\{\sum_{j=1}^N A_j^3,\ \calH_k\Bigr\} = 0. 
\eeq
To obtain a commuting set of flows, therefore, 
one has to impose the constraint 
\beq
    \sum_{j=1}^N A_j^3 = 0. 
    \label{eq:A-constraint}
\eeq
Note that $\sum_{j=1}^N A_j^3$ is 
an infinitesimal generator of the diagonal 
gauge transformations 
\beq
  A_j \mapsto g^{-1}A_jg, \quad 
  g = \left(\begin{array}{cc} 
      e^\gamma & 0 \\
      0 & -e^\gamma 
      \end{array}\right). 
  \label{eq:diagonal-gt}
\eeq

Because of the presence of the constraint 
(\ref{eq:A-constraint}), as Korotkin and 
Samtleben \cite{bib:Ko-Sa} pointed out, 
one has to modify the naive Hamiltonians $\calH_j$ 
by a term proportional to the left hand side of 
the constraint as 
\beq
    \widetilde{\calH}_j = \calH_j - 2p_j\sum_{k=1}^N A_k^3. 
\eeq
The multipliers $p_j$, $j = 1,\ldots,N$, are assumed 
to satisfy the relation 
\beq
    \sum_{j=1}^N p_j = p, 
    \label{eq:p-constraint}
\eeq
which ensures the consistency of Lax and zero-curvature 
equations that we shall derive later on.   
A set of commuting flows are now defined on the reduced 
phase space by the canonical equations 
\beq
    \frac{\rd q}{\rd t_j} = \{q, \widetilde{\calH}_j\}, \quad 
    \frac{\rd p}{\rd t_j} = \{p, \widetilde{\calH}_j\}, \quad 
    \frac{\rd A_k}{\rd t_j} = \{A_k, \widetilde{\calH}_j\}. 
    \label{eq:qpA-system}
\eeq
The Poisson brackets on the right hand side are understood 
to be calculated as 
\beq
  \{F, \widetilde{\calH}_j\} 
  = \{F, \calH_j\} - 2p_j\{F,\; \sum_{j=1}^N A_j\}. 
\eeq
Namely, we first calculate the Poisson bracket in 
the unconstrained variables, then impose the constraint.  
Since the term containing $\{F, p_j\}$ disappears upon 
imposing the constraint, we leave the Poisson brackets 
with $p_j$'s undefined.  More explicitly, the equations 
of motion read as follows. 
\begin{enumerate}
\item Equations of motion of $q$ and $p$: 
\beq
  \frac{\rd q}{\rd t_j} &=& 2A_j^3, \nonumber \\
  \frac{\rd p}{\rd t_j} &=& 
    - \sum_{k\not= j} \phi_u(q,t_j-t_k)A_j^{+}A_l^{-} 
    + \sum_{k\not= j} \phi_u(-q,t_j-t_k)A_j^{-}A_k^{+}.  
  \label{eq:eom-qp}
\eeq
\item Equations of motion of $A_k^{\pm}$: 
\beq
  \frac{\rd A_k^{\pm}}{\rd t_j} &=& 
    \mp 2\zeta(t_j-t_k)A_j^3A_k^{\pm} 
    \pm 2\phi(\pm q,t_j-t_k)A_j^{\pm}A_k^3 
    \pm 2p_jA_k^{\pm} 
    \quad (j\not= k), \nonumber \\
  \frac{\rd A_k^{\pm}}{\rd t_k} &=& 
    \mp 2\sum_{\ell\not= k}\zeta(t_k-t_\ell)A_k^{\pm}A_\ell^3 
    \pm 2\sum_{\ell\not= k}\phi(\mp q,t_k-t_\ell)A_k^3A_\ell^{\pm} 
    \mp 2\sum_{\ell\not= k}p_\ell A_k^{\pm}. 
  \label{eq:eom-Apm}
\eeq
\item Equations of motion of $A_k^3$: 
\beq
  \frac{\rd A_k^3}{\rd t_j} &=& 
    - \phi(q,t_j-t_k)A_j^{+}A_k^{-} 
    + \phi(-q,t_j-t_k)A_j^{-}A_k^{+} 
    \quad (j\not= k), \nonumber \\
  \frac{\rd A_k^3}{\rd t_k} &=& 
    \sum_{\ell\not= k}\phi(q,t_k-t_\ell)A_k^{+}A_\ell^{-} 
    - \sum_{\ell\not= k}\phi(-q,t_k-t_\ell)A_k^{-}A_\ell^{+}. 
  \label{eq:eom-A3}
\eeq
\end{enumerate}
In particular, the sum of these flows turns out to be trivial: 
\beq
  \sum_{j=1}^N \frac{\rd q}{\rd t_j} = 0, \quad 
  \sum_{j=1}^N \frac{\rd p}{\rd t_j} = 0, \quad 
  \sum_{j=1}^N \frac{\rd A_k}{\rd t_j} = 0. 
  \label{eq:sum-of-flows-trivial}
\eeq
The $t$-dependence of $p_j$'s cannot be determined 
in this way;  we shall derive a set of differential 
equations for $p_j$'s later on.

\subsection{Calculating $\{L(z),\widetilde{\calH}_j\}$}

As an intermediate step towards Lax equations, 
we now consider the Poisson bracket 
\beq
    \{L(z), \widetilde{\calH}_j\} 
    = \{L(z), \calH_j\} - 2p_j\{L(z),\sum_{k=1}^N A_k^3\} 
\eeq
of $L(z)$ with the modified Hamiltonians $\widetilde{H}_j$. 
A clue is the general formula 
\beq
    \Bigl\{L(z), \ \frac{1}{n}\Tr L(w)^n\Bigr\} 
    = \Tr{}_2\Bigl(\{L(z) \otcomma L(w)\} 
        I \otimes L(w)^{n-1}\Bigr), 
\eeq
where $\Tr{}_2$ denotes the trace over the second 
component of the tensor product: 
\beqnn
    \Tr{}_2 \Bigl(\sum_{a,b,c,d}X_{abcd}
      E_{ab} \otimes E_{cd}\Bigr) 
    = \sum_{a,b} \Bigl(\sum_c X_{abcc}\Bigr) E_{ab}. 
\eeqnn
Plugging the Poisson commutation relation 
(\ref{eq:PCR-for-L}) into this formula, 
one obtains the identity 
\beq
    \Bigl\{L(z), \ \frac{1}{2}\Tr L(w)^2\Bigr\} 
    &=& \Bigl[L(z),\ \Tr{}_2(r(z-w)I \otimes L(w))\Bigr] 
    \nonumber \\
    && \mbox{}
      + 2 \Tr{}_2\Bigl(\frac{\rd r(z-w)}{\rd q}
          I \otimes L(w)\Bigr) \sum_{j=1}^N A_j^3. 
\eeq
Extracting the residues at $w = t_j$ yields 
the Poisson bracket $\{L(z),\calH_j\}$.  Note that 
the second term on the right hand side 
disappears upon imposing the constraint.  Since 
\beqnn
    \Res_{w=t_j} \Tr{}_2(r(z-w)I \otimes L(w))
    = \left(\begin{array}{cc}
      \zeta(z-t_j)A_j^3 & \phi(q,z-t_j)A_j^{-} \\
      \phi(-q,z-t_j)A_j^{+} & - \zeta(z-t_j)A_j^3
      \end{array}\right), 
\eeqnn 
the Poisson bracket with $\calH_j$ eventually 
takes the form 
\beqnn
    \{L(z), \calH_j\} = [L(z), M_j(z)], 
\eeqnn
where 
\beq
    M_j(z) =  
        \left(\begin{array}{cc}
        \zeta(z-t_j)A_j^3 & \phi(q,z-t_j)A_j^{-} \\
        \phi(-q,z-t_j)A_j^{+} & - \zeta(z-t_j)A_j^3
        \end{array}\right). 
\eeq
On the other hand, the Poisson bracket with 
$\sum_{k=1}^N A_k^3$ can be readily calculated as 
\beqnn
  \{L(z), \sum_{k=1}^N A_k^3\} 
  &=& \sum_{k=1}^N 
      \left(\begin{array}{cc} 
      0  & \phi(q, z-t_k)A_k^{-} \\
      \phi(-q, z-t_k)A_k^{+} & 0 
      \end{array}\right) \nonumber \\
  &=& [L(z), \diag(-1/2, 1/2)]. 
\eeqnn
One thus finds that 
\beq
  \{L(z), \widetilde{\calH}_j\} = [L(z), \widetilde{M}_j(z)], 
\eeq
where 
\beq
  \widetilde{M}_j(z) = M_j(z) + \diag(p_j,-p_j). 
\eeq

\subsection{Lax equations}

We are now ready to derive the Lax equations.  

By the Leibniz rule, the $t$-derivatives of the matrix elements 
of $L(z)$ can be written as 
\beqnn
    \frac{\rd L_{12}(z)}{\rd t_j} 
    = \sum_{k=1}^N 
        \Bigl(\phi_u(q,z-t_k)\frac{\rd q}{\rd t_j}A_k^{-}  
          + \phi(q,z-t_k)\frac{\rd A_k^{-}}{\rd t_j}\Bigr) 
      - \phi'(q,z-t_j) A_j^{-}, 
    \nonumber \\
    \frac{\rd L_{21}(z)}{\rd t_j} 
    = \sum_{k=1}^N 
        \Bigl(-\phi_u(-q,z-t_j)\frac{\rd q}{\rd t_j}A_k^{+} 
          + \phi(-q,z-t_j)\frac{\rd A_k^{+}}{\rd t_j}\Bigr) 
      - \phi'(-q,z-t_j)A_j^{+}, 
\eeqnn
and 
\beqnn
  \frac{\rd L_{11}(z)}{\rd t_j} 
  = - \frac{\rd L_{22}(z)}{\rd t_j} 
  = \frac{\rd p}{\rd t_j} 
    + \sum_{k=1}^N \zeta(z-t_k)\frac{\rd A_k^3}{\rd t_j} 
    - \zeta'(z - t_j)A_j^3, 
\eeqnn
where 
\beqnn
    \phi'(u,z) = \frac{\rd \phi(q,z)}{\rd z} 
    = - \phi(u,z)(\zeta(u - z) + \zeta(z)). 
\eeqnn
Notice here that the terms containing $\phi'(\pm q,z - t_j)$ 
and $\zeta(z - t_j)$ coincide with the matrix elements 
of $-\rd\widetilde{M}_j(z)/\rd z$.  On the other hand, 
the $t_j$-derivatives of the dynamical variables 
$q,p,A_k$ can be expressed as the Poisson bracket 
with $\widetilde{\calH}_j$.  Consequently, 
\beqnn
    \frac{\rd L(z)}{\rd t_j} 
    = \{L(z),\calH_j\} - \frac{\rd \widetilde{M}_j(z)}{\rd z}. 
\eeqnn
Combining this with the foregoing calculation of 
$\{L(z),\calH_j\}$, we eventually obtain 
the Lax equations 
\beq
    \frac{\rd L(z)}{\rd t_j} 
    = [L(z), \widetilde{M}_j(z)] 
      - \frac{\rd \widetilde{M}_j(z)}{\rd z} 
    \label{eq:Lax} 
\eeq
of the isomonodromic type.   Though we omit details, 
one can conversely derive the equations of motion of 
$q,p$ and $A_k$'s from these Lax equations.  

As a remark, let us mention that these Lax equations 
are consistent with (\ref{eq:sum-of-flows-trivial}).  
This is a consequence of the linear relation 
(\ref{eq:p-constraint}) among $p_j$'s and $p$.  
One can derive, from this linear relation, 
the linear relation 
\beq
  \sum_{j=1}^N \widetilde{M}_j(z) = L (z) 
\eeq 
among the matrices $\widetilde{M}_j(z)$ and $L(z)$. 
The sum of the $N$ Lax equations thereby yields 
the relation 
\beq
  \sum_{j=1}^N \frac{\rd L(z)}{\rd t_j} 
  = - \frac{\rd L(z)}{\rd z},  
\eeq
which is a restatement of 
(\ref{eq:sum-of-flows-trivial}).

\subsection{Zero-curvature equations}

The Lax equations are self-consistent 
(i.e., define commutating flows) if and only if 
the zero-curvature equations 
\beq
  \frac{\rd\widetilde{M}_k(z)}{\rd t_j} 
  - \frac{\rd\widetilde{M}_j(z)}{\rd t_k} 
  + [\widetilde{M}_j(z), \widetilde{M}_k(z)] = 0 
  \label{eq:zc} 
\eeq
are satisfied.  As we show below, these equations 
give a set of differential equations for $p_j$'s.  

The differential equations for $p_j$'s are obtained 
from the  the diagonal part of the zero-curvature 
equations.  We have only to consider the upper left 
component, because all matrices in the zero-curvature 
equations are trace-free.  The diagonal part 
$[\widetilde{M}_j(z),\widetilde{M}_k(z)]_{11} 
= - [\widetilde{M}_j(z),\widetilde{M}_k(z)]_{22}$ 
of the commutator can be rewritten as 
\beqnn
  [\widetilde{M}_j(z),\widetilde{M}_k(z)]_{11} 
  &=& \phi(q,z - t_j)\phi(-q,z - t_k)A_j^{-}A_k^{+} 
    - \phi(q,z - t_k)\phi(-q,z - t_j)A_k^{-}A_j^{+} \\
  &=& - \Bigl(\phi(q,t_k - t_j)(\zeta(z - t_j) - \zeta(z - t_k)) 
          + \phi_u(q,t_k - t_j)\Bigr)A_j^{-}A_k^{+} \\
  &&\mbox{} 
      + \Bigl(\phi(q,t_j - t_k)(\zeta(z - t_k) - \zeta(z - t_j)) 
          + \phi_u(q,t_j - t_k)\Bigr)A_k^{-}A_j^{+}. 
\eeqnn
The functional identity (\ref{eq:phi(u,z)phi(-u,z)}) 
has been used in the last stage.  The $z$-dependent pieces 
are thus a linear combination of $\zeta(z - t_j)$ and 
$\zeta(z - t_k)$.  On the other hand, the derivative part 
$\rd\widetilde{M}_{k,11}(z)/\rd t_j - \rd\widetilde{M}_{j,11}(z)/\rd t_k$ 
of the zero-curvature equation can be expressed as 
\beqnn
    \frac{\rd\widetilde{M}_{k,11}(z)}{\rd t_j} 
    - \frac{\rd\widetilde{M}_{j,11}(z)}{\rd t_k} 
  = \frac{\rd p_k}{\rd t_j} + \zeta(z - t_k)\frac{\rd A_k^3}{\rd t_j} 
    - \frac{\rd p_j}{\rd t_k} - \zeta(z - t_j)\frac{\rd A_j^3}{\rd t_k}. 
\eeqnn
Upon summation with the foregoing expression of the commutator part, 
the $z$-dependent pieces $\zeta(z - t_j)(\cdots) + \zeta(z - t_k)(\cdots)$ 
turn out to cancel out as a consequence of the equations of motion of 
$A_j^3$ and $A_k^3$.  One is thus left with the differential equations 
\beq
  \frac{\rd p_k}{\rd t_j} - \frac{\rd p_j}{\rd t_k} 
  + \phi_u(q,t_j - t_k)A_j^{+}A_k^{-}
  - \phi_u(q,t_k - t_j)A_j^{-}A_k^{+} 
  = 0 
  \label{eq:zc-pj}
\eeq
for $p_j$'s.  

One can examine the off-diagonal part of the zero-curvature equations 
in much the same way.  This however leads to no new equations 
(see Appendix C).  Namely, all equations are satisfied under 
the equations of motion of $q,p$ and $A_j$'s.  

The final problem is the existence of a solution of 
(\ref{eq:zc-pj}).  To this end, it is convenient 
to convert (\ref{eq:zc-pj}) to an exterior differential 
equation. Let $\theta$ denote the one-form 
\beq
  \theta = \sum_{j=1}^N p_j dt_j. 
\eeq
(\ref{eq:zc-pj}) thereby turns into 
the exterior differential equation 
\beq
  d\theta = \omega, 
\eeq 
where 
\beq
  \omega 
  &=& - \frac{1}{2}\sum_{j,k=1}^N \phi_u(q,t_j-t_k)A_j^{+}A_k^{-} 
      + \frac{1}{2}\sum_{j,k=1}^N \phi_u(q,t_k-t_j)A_j^{-}A_k^{+} 
  \nonumber \\
  &=& \sum_{j,k=1}^N 
        \phi(q,t_j-t_k)(\zeta(t_j-t_k-q) + \zeta(q))A_j^{+}A_k^{-}. 
\eeq
Consequently, the Frobenius integrability of 
(\ref{eq:zc-pj}) is equivalent to the closedness 
\beq
  d\omega = 0 
\eeq
of $\omega$.  Actually, this integrability condition 
turns out to be satisfied under the equations of 
motion of $q,p$ and $A_j$ (see Appendix D).

\section{Spectral curve and Darboux coordinates}

\subsection{Spectral curve}

The spectral curve is defined by the eigenvalue equation 
\beq
    \det(wI - L(z)) = 0. 
\eeq
Since $L(z)$ is trace-free, the left hand side can be 
rewritten as 
\beq
    \det(wI - L(z)) = w^2 + \det L(z) 
                    = w^2 - \frac{1}{2}\Tr L(z)^2. 
\eeq

Under the constraint (\ref{eq:A-constraint}), 
the matrix elements of $L(z)$ enjoy the following 
quasi-periodicity along the period lattice of the torus: 
\beq
    L(z + 2m\omega_1 + 2n\omega_3) 
    = e^{-(m\eta_1 + n\eta_3) Q} L(z) 
      e^{(m\eta_1 + n\eta_3) Q}, 
\eeq
where $Q$ is the diagonal matrix $Q = \diag(q,-q)$, 
and $\eta_1$ and $\eta_3$ are the values of $\zeta(z)$ 
at $z = \omega_1,\omega_3$.  The quasi-periodicity of 
$L(z)$ is a consequence of the quasi-periodicity of 
$\zeta(z)$ and $\phi(u,z)$, 
\beq
    \zeta(z + 2m\omega_1 + 2n\omega_3) 
    &=& \zeta(z) + 2m\eta_1 + 2n\omega_3, \\
    \phi(u, z + 2m\omega_1 + 2m\omega_3) 
    &=& e^{-2m\eta_1-2n\eta_3}\phi(u,z), 
\eeq
which are easy to confirm from the property of 
the sigma function. 

The quasi-periodicity of $L(z)$, in particular, 
implies the double periodicity of $\Tr L(z)^2/2$, 
which thereby becomes an elliptic function with 
poles at $z = t_1,\ldots,t_N$.  Since 
\beqnn
    L(z) = \frac{A_j}{z-t_j} + O(1)  
\eeqnn
as $z \to t_j$, this elliptic function has a double pole 
at $z = t_j$ with the leading coefficient equal to the 
quadratic Casimir 
\beq 
    C_j = \frac{1}{2}\Tr A_j^2 = \frac{\theta_j^2}{4}
\eeq
of $A_j$.  The residue is nothing but the Hamiltonian 
$\calH_j$.  Thus $\Tr L(z)^2/2$ can be expressed as 
\beq
   \frac{1}{2}\Tr L(z)^2 
   = \sum_{j=1}^N C_j \wp(z-t_j) 
   + \sum_{j=1}^N \calH_j \zeta(z-t_j) 
   + \calH_0, 
\eeq
where $\calH_0$ is a constant term (which however 
depends on $\omega_1$ and $\omega_3$).  Also note 
that the Hamiltonians are not linearly independent, 
but obey the linear constraint 
\beq
    \sum_{j=1}^N \calH_j = 0.  
    \label{eq:calH-constraint}
\eeq
This is a consequence of the the double periodicity
of $\Tr L(z)^2/2$. 

The spectral curve thus turns out to be a double 
covering of the torus. The branch points are 
located above the (possibly multiple) $2N$ zeros 
of $\Tr L(z)^2/2$.  If these zeros are all simple, 
the genus of the spectral curve is equal to $N + 1$.  
The spectral curve is time-dependent because of 
the extra term $\rd M_j(z)/\rd z$ on the right 
hand side of the Lax equation.

\subsection{Spectral Darboux coordinates}

The construction of spectral Darboux coordinates is 
parallel to the case of the rational (and isospectral) 
model \cite{bib:Br}.  The ``coordinate part''
$\lambda_1,\ldots,\lambda_N$ are defined as 
the $N$ zeros (modulo the period lattice) 
of $L_{12}(z)$, 
\beq
    L_{12}(\lambda_j) = 0, 
\eeq
and the ``momentum part'' $\mu_1,\ldots,\mu_N$ are 
defined to be the value of $L_{11}(z)$ at these points, 
\beq
    \mu_j = L_{11}(\lambda_j) 
          = p + \sum_{k=1}^N \zeta(\lambda_j - t_k)A_k^3. 
    \label{eq:mu-def}
\eeq
In order to avoid a delicate problem, we assume 
throughout the following consideration that 
$\lambda_j \not= \lambda_k$ if $j \not= k$. 
It is easy to see that $(\lambda_j,\mu_j)$ sits 
on the spectral curve; $L(\lambda_j)$ takes the 
triangular form 
\beqnn
    L(\lambda_j) 
    = \left(\begin{array}{cc}
      \mu_k & 0 \\
      L_{21}(\lambda_j) & - \mu_k
      \end{array}\right), 
\eeqnn
which implies that $\pm\mu_j$ are eigenvalues 
of $L(\lambda_j)$. 

The $\lambda_j$'s are constrained by a linear relation.  
To see this, let us note that $L_{12}(z)$ can be 
factorized as 
\beq
    L_{12}(z) 
    = \kappa \frac{\prod_{j=1}^N\sigma(z - \lambda_j)}
                  {\prod_{j=1}^N\sigma(z - t_j)}, 
    \label{eq:L12-factor}
\eeq
where $\kappa$ is a constant that does not depend on $z$.  
The quasi-periodicity 
\beqnn
    L_{12}(z + 2m\omega_1 + 2n\omega_3) 
    = e^{-(2m\eta_1 + 2n\eta_3)q}L_{12}(z) 
\eeqnn
of $L_{12}(z)$ implies that its zeros 
$\lambda_1,\ldots,\lambda_N$ are constrained as 
\beq
    \sum_{j=1}^N \lambda_j - \sum_{j=1}^N t_j \equiv q 
    \quad \mbox{mod} \ 2\omega_1\ZZ + 2\omega_3\ZZ. 
\eeq
Since each $\lambda_j$ is defined only up to 
a difference by an element of the period lattice, 
let us redefine $\lambda_j$'s, if necessary, such 
that this holds without ``mod $2\omega_1\ZZ + 2\omega_3\ZZ$'': 
\beq
    \sum_{j=1}^N \lambda_j - \sum_{j=1}^N t_j = q.  
    \label{eq:q-constraint}
\eeq
Of course this will be valid only for a {\it local} 
study of the system; this naive prescription has to be 
modified if one considers a global problem.  

Let us note here that the coefficient $\kappa$ transforms as 
$\kappa \to e^{2\gamma}\kappa$ under diagonal gauge transformation 
(\ref{eq:diagonal-gt}).  One can thereby adjust $\kappa$ 
to any non-zero value, e.g., 
\beq
  \kappa = 1. 
  \label{eq:gauge-fixing}
\eeq
In other words, $\kappa$ is not a true dynamical degree 
of freedom.

\subsection{Time-dependent canonical transformation}

In order to prove the canonicity of these variables 
$\lambda_j,\mu_j$, we now restrict the spin variables onto 
the direct product $\calO_1 \times \cdots \times \calO_N$ 
of the symplectic leaves and use the parametrization 
(\ref{eq:A-by-xxi}) by $(x_j,\xi_j)$.  Moreover, 
we tentatively relax the constraint (\ref{eq:A-constraint}), 
which now takes the form 
\beq
    \sum_{j=1}^N x_j \xi_j = 0, 
    \label{eq:xxi-constraint}
\eeq
and restore it in the final stage.   

The factorization relation (\ref{eq:L12-factor}) 
of $L_{12}(z)$ now reads 
\beq
    \frac{1}{2} \sum_{j=1}^N \phi(q, z-t_j)x_j^2 
    = \kappa \frac{Q(z)}{P(z)} 
\eeq
where we have introduced the two functions 
\beq
    Q(z) = \prod_{j=1}^N \sigma(z - \lambda_j), \quad 
    P(z) = \prod_{j=1}^N \sigma(z - t_j). 
\eeq
This reduces to the relations 
\beq
    \frac{1}{2}x_j^2 
    = \kappa\frac{Q(t_j)}{P'(t_j)} 
    = \kappa\frac{\prod_{k=1}^N \sigma(t_j - \lambda_k)}
              {\prod_{k\not= j} \sigma(t_j - t_k)} 
\eeq
of the residues of both hand sides at $z = t_j$.  
These relations show how the old variables $x_j$ 
are connected with the new variables $\lambda_j$ 
(and $\kappa$).  By logarithmic differentiation, 
these relations can be further converted to the 
linear relations 
\beq
    2d\log x_j 
    &=& d\log\kappa 
    + \sum_{k=1}^N d\log\sigma(t_j - \lambda_k) 
    - \sum_{k\not= j} d\log\sigma(t_j - t_k) 
    \nonumber \\
    &=& d\log\kappa 
    + \sum_{k=1}^N \zeta(t_j - \lambda_k) (dt_j - d\lambda_k) 
    - \sum_{k\not= j} \zeta(t_j - t_k) (dt_j - dt_k). 
\eeq
of differential forms.  

Our goal is to derive a relation between the 
canonical one-forms $\sum_{j=1}^N \xi_jdx_j + pdq$ and 
$\sum_{j=1}^N \mu_jd\lambda_j$.  We first multiply 
the both hand sides of the last relation by 
$x_j\xi_j/2$, sum over $j = 1,\ldots,N$, 
and add $pdq$ to both hand sides.  We then obtain 
the linear relation 
\beqnn
    \sum_{j=1}^N \xi_jdx_j + pdq 
    &=& \frac{1}{2}\sum_{j=1}^N x_j\xi_jd\log\kappa 
      + \frac{1}{2}\sum_{j,k=1}^N x_j\xi_j 
          \zeta(t_j - \lambda_k)(dt_j - d\lambda_k) 
    \nonumber \\
    &&  \mbox{} 
      - \frac{1}{2}\sum_{j\not= k}x_j\xi_j 
          \zeta(t_j - t_k)(dt_j - dt_k) 
      + pdq. 
\eeqnn 
On the other hand, by differentiating 
(\ref{eq:q-constraint}), we have the relation 
\beqnn
    dq = \sum_{j=1}^N d\lambda_j - \sum_{j=1}^N dt_j, 
\eeqnn
which we can use to eliminate the differential $dq$ 
on the right hand side of the foregoing linear 
relation of one-forms.  The right hand side thereby 
becomes a linear combination of $d\log\kappa$, 
$d\lambda_j$'s and $dt_j$'s,  and the coefficient 
of $d\lambda_j$ turns out to be equal to $\mu_j$ by 
(\ref{eq:mu-def}).  We thus eventually find that 
\beq
    \sum_{j=1}^N \xi_j dx_j + pdq 
    &=& \frac{1}{2}\sum_{j=1}^N x_j\xi_j d\log\kappa 
      + \sum_{j=1}^N \mu_j d\lambda_j 
    \nonumber \\
    &&  \mbox{}
      - p\sum_{j=1}^N dt_j 
      + \frac{1}{2}\sum_{j,k=1}^N x_j\xi_j 
          \zeta(t_j - \lambda_k)dt_j 
    \nonumber \\
    &&  \mbox{}
      - \frac{1}{2}\sum_{j\not= k} x_j\xi_j 
          \zeta(t_j - t_k)(dt_j - dt_k). 
\eeq

The last equation shows that $\lambda_j$ and $\mu_j$ 
are Darboux coordinates of the canonical one-form 
$\sum_{j=1}^N \xi_j dx_j + pdq$, and that $\log\kappa$ 
is a conjugate variable of the left hand side 
of constraint (\ref{eq:xxi-constraint}). 
This interpretation is fully parallel to 
the spectral description of rational isospectral 
systems \cite{bib:AHH,bib:Ha-dual,bib:Ha-Wi}.  

An essential difference lies in the fact that 
the time variables explicitly enter the relation 
between the two canonical one-forms.  This means 
that the spectral Darboux coordinates are connected 
with the old variables $(x_j,\xi_j,q,p)$ by a 
{\it time-dependent} canonical transformation.  
Accordingly, the Hamiltonians (which is denoted 
by $H_j$ in the following) in the spectral 
Darboux coordinates differ from the Hamiltonians 
$\calH_j$ in the old variables $(x_j,\xi_j,q,p)$.  
Their relation is to be determined by the 
fundamental formula 
\beq
    \sum_{j=1}^N \xi_jdx_j + pdq - \sum_{j=1}^N \calH_jdt_j 
  = \sum_{j=1}^N \mu_jd\lambda_j - \sum_{j=1}^N H_jdt_j, 
\eeq
in which we have imposed the constraint 
(\ref{eq:xxi-constraint}).  More explicitly, 
the new Hamiltonians are defined as 
\beq
    H_j = \calH_j + p 
      - \frac{1}{2}x_j\xi_j
          \Bigl(\sum_{k=1}^N \zeta(t_j - \lambda_k) 
            - \sum_{k\not= j}\zeta(t_j - t_k)\Bigr) 
      + \frac{1}{2}\sum_{k\not= j}\zeta(t_j -t_k)x_k\xi_k. 
    \label{eq:H-def}
\eeq
The goal of the next section is to rewrite the right 
hand side in terms of the spectral Darboux coordinates.

\section{Hamiltonian system in spectral Darboux coordinates}

\subsection{Linear equations characterizing Hamiltonians}

Let us recall that the pairs $(\lambda_j,\mu_j)$ 
of the spectral Darboux coordinates all sit on 
the spectral curve.  Therefore the equations 
\beqnn
    \mu_k^2 = \sum_{j=1}^N C_j\wp(\lambda_k - t_j) 
            + \sum_{j=1}^N \calH_j\zeta(\lambda_k - t_j) 
            + \calH_0 
\eeqnn
are satisfied for $k = 1,\ldots,N$.  These equations, 
along with the linear constraint (\ref{eq:calH-constraint}), 
may be thought of as a system of linear equations 
that determine $\calH_j$'s.  In fact, as we shall 
discuss afterwards, these linear equations can be solved 
for $\calH_j$'s, which thereby becomes an explicit 
function of the spectral Darboux coordinates (and of 
the time variables).  

If the system in consideration were an isospectral 
system (such as the Moser system or the usual 
Calogero-Gaudin system), the time variables would 
not appear here explicitly, and $\calH_j$'s would be 
the Hamiltonians that we have sought.  The only problem 
would have been to solve the foregoing linear equations 
for $\calH_j$.  This is, actually, what Brzezi\'nski 
\cite{bib:Br} and Enriquez et al. \cite{bib:En-Fe-Ru} 
did in their work on separation of variables of 
the $\rmSU(2)$ Calogero-Gaudin system.  

In the present case, the true Hamiltonians 
are {\it not} $\calH_j$'s {\it but} $H_j$'s.  
We have to rewrite the extra terms on 
the right hand side of (\ref{eq:H-def}), 
too, as a function of the spectral Darboux 
coordinates. This is another problem that 
we have to solve.  

To this end, let us note that the defining 
equation (\ref{eq:mu-def}) of $\mu_j$, 
which can be rewritten as 
\beqnn
    \mu_k 
  = \sum_{j=1}^N\zeta(\lambda_k - t_j)\frac{x_j\xi_j}{2} 
    + p, 
\eeqnn
may be thought of as a system of linear equations 
for $p$ and $x_j\xi_j/2$.  If one can solve these 
equations for $p$ and $x_j\xi_j/2$, the solution 
should be an expression of $p$ and $x_j\xi_j/2$ 
as a function of the spectral Darboux coordinates. 
Remarkably, these linear equations have the same 
coefficients as the foregoing linear equations 
for $\calH_0$ and $\calH_j$.  Moreover, $x_j\xi_j/2$ 
obey the linear constraint (\ref{eq:xxi-constraint}), 
in perfect analogy with the linear constraint 
(\ref{eq:calH-constraint}) for $\calH_j$.  

Thus the two problems, one for $\calH_j$ and 
the other for the extra terms in (\ref{eq:H-def}), 
can be reduced to a single problem, namely, 
solving a system of linear equations of the form 
\beq
  \begin{array}{rl}
    \displaystyle\sum_{j=1}^N \zeta(\lambda_k - t_j)X_j + X_0 = b_k 
    & (k=1,\ldots,N), \\
    \displaystyle\sum_{j=1}^N X_j = 0 & 
  \end{array}
    \label{eq:lineq}
\eeq
As we shall show below, this system of linear equations 
has a unique and explicit solution.

\subsection{Solution of linear equations}

We assume that $q \not\equiv 0 \ 
\mbox{mod}\ 2\omega_1\ZZ + 2\omega_3\ZZ$ or, 
equivalently, $\sigma(q) \not= 0$.  The following 
ensures the uniqueness of solution of (\ref{eq:lineq}). 

\begin{lemma} 
If $b_j = \cdots = b_N = 0$, then 
$X_0 = X_1 = \cdots = X_N = 0$.  
\end{lemma}

\proof 
Consider the function 
\beqnn
    f(z) = \sum_{j=1}^N \zeta(z - t_j)X_j + X_0. 
\eeqnn
The first $N$ equations of (\ref{eq:lineq}) imply that 
$f(z)$ has zeros at $z =\lambda_1,\ldots,\lambda_N$. 
The remaining one ensures that $f(z)$ is a doubly 
periodic meromorphic function on the $z$-plane.  
All possible poles are obviously simple and confined 
to $z = t_1,\ldots,t_N$ and their translations by 
the period lattice.  Therefore, if $f(z)$ is 
not identically zero, the zeros $\lambda_j$ 
and the poles $t_j$ are constrained as 
\beqnn
    \sum_{j=1}^N \lambda_j- \sum_{j=1}^N t_j \equiv 0 
    \quad \mbox{mod}\ 2\omega_1\ZZ + 2\omega_3\ZZ, 
\eeqnn
but this contradicts the assumption that 
$q \not\equiv 0$; recall the constraint 
(\ref{eq:q-constraint}).  Thus $f(z)$ is 
identically zero, and all the coefficients 
$X_0,X_1,\cdots,X_N$ have to be zero. \qed

Having proven the uniqueness, the problem is 
to find a solution by any means.  This can be 
done with the aid of an elliptic analogue of 
Lagrange's interpolation formula (see Appendix A). 

\begin{lemma}
A solution of (\ref{eq:lineq}) is given by 
\beq
  X_j &=& 
  \sum_{k=1}^N 
  \frac{Q(t_j)P(\lambda_k)\sigma(t_j-\lambda_k+q)b_k}
       {P'(t_j)Q'(\lambda_k)\sigma(t_j-\lambda_k)\sigma(q)},
  \label{eq:lineq-sol1}\\
  X_0 &=& 
  - \sum_{j,k=1}^N 
    \frac{Q(t_j)P(\lambda_k)\sigma(t_j-\lambda_k+q)
          \zeta(\lambda_k-t_j-q)b_k}
         {P'(t_j)Q'(\lambda_k)\sigma(t_j-\lambda_k)\sigma(q)}. 
  \label{eq:lineq-sol2}
\eeq
\end{lemma}

\proof
We have only to confirm that these $X_j$ and $X_0$ 
do satisfy (\ref{eq:lineq}).  The last equation 
of (\ref{eq:lineq}) is indeed satisfied as 
(\ref{eq:sum-to-zero}) shows. As regards the 
other equations of (\ref{eq:lineq}), the main task 
is to calculate 
\beqnn
  \sum_{j=1}^N \zeta(\lambda_l-t_j)X_j 
  = \sum_{k=1}^N \Bigl( \sum_{j=1}^N 
        \frac{Q(t_j)\sigma(t_j-\lambda_k+q)}
             {P'(t_j)\sigma(t_j-\lambda_k)}
        \zeta(\lambda_l - t_j) \Bigr) 
    \frac{P(\lambda_k)b_k}
         {Q'(\lambda_k)\sigma(q)}. 
\eeqnn
We can use the two identities (\ref{eq:interpol-1}) 
and (\ref{eq:interpol-2}) to rewrite the sum over 
$j$ on the right hand side, and find that 
\beqnn
  \sum_{j=1}^N \zeta(\lambda_l-t_j)X_j 
  &=& \sum_{k,j=1}^N 
      \frac{Q(t_j)P(\lambda_k)\sigma(t_j-\lambda_k+q)
            \zeta(\lambda_k-t_j-q)b_k}
           {P'(t_j)Q'(\lambda_k)\sigma(t_j-\lambda_k)\sigma(q)} 
    + b_l \\
  &=& - X_0 + b_l, 
\eeqnn
which is nothing but the first $N$ equations 
of (\ref{eq:lineq}).  
\qed

\subsection{Writing $H_j$ explicitly}

Let us apply the foregoing formulae 
(\ref{eq:lineq-sol1}) and (\ref{eq:lineq-sol2}) 
of solution of (\ref{eq:lineq}) to the problem 
of deriving an explicit form of $H_j$ as a function 
of the spectral Darboux coordinates.  

If we use the formulae to the case where 
\beqnn
    X_j = \calH_j, \quad 
    X_0 = \calH_0, \quad 
    b_k = \mu_k^2 - \sum_{l=1}^N C_l\wp(\lambda_k - t_l), 
\eeqnn
we find the following expression of $\calH_j$ and $\calH_0$: 
\beq
  \calH_j &=& 
  \sum_{j=1}^N 
  \frac{Q(t_j)P(\lambda_k)\sigma(t_j-\lambda_k+q)}
       {P'(t_j)Q'(\lambda_k)\sigma(t_j-\lambda_k)\sigma(q)} 
  \Bigl(\mu_k^2 - \sum_{l=1}^N C_l\wp(\lambda_k - t_l)\Bigr), 
  \\
  \calH_0 &=& 
  - \sum_{j,k=1}^N 
    \frac{Q(t_j)P(\lambda_k)\sigma(t_j-\lambda_k+q)
          \zeta(\lambda_k-t_j-q)}
         {P'(t_j)Q'(\lambda_k)\sigma(t_j-\lambda_k)\sigma(q)} 
    \Bigl(\mu_k^2 - \sum_{l=1}^N C_l\wp(\lambda_k - t_l)\Bigr). 
\eeq
Similarly, if we use the formulae in the case where 
\beqnn
    X_j = \frac{1}{2}x_j\xi_j, \quad 
    X_0 = p, \quad 
    b_k = \mu_k. 
\eeqnn
we find the following expression of $x_j\xi_j/2$ and $p$ as 
a function of the spectral Darboux coordinates: 
\beq
  \frac{1}{2}x_j\xi_j &=& 
  \sum_{k=1}^N 
  \frac{Q(t_j)P(\lambda_k)\sigma(t_j-\lambda_k+q)\mu_k}
  {P'(t_j)Q'(\lambda_k)\sigma(t_j-\lambda_k)\sigma(q)}, 
  \label{eq:xxi-by-lamu}
  \\
  p &=& 
  - \sum_{j,k=1}^N 
    \frac{Q(t_j)P(\lambda_k)\sigma(t_j-\lambda_k+q) 
          \zeta(\lambda_k-t_j-q)\mu_k} 
    {P'(t_j)Q'(\lambda_k)\sigma(t_j-\lambda_k)\sigma(q)}. 
  \label{eq:p-by-lamu}
\eeq
Thus we have been able to rewrite each term 
on the right hand side of (\ref{eq:H-def}) to 
an explicit function of the spectral Darboux 
coordinates.  

Although the extra terms on the right hand side 
of (\ref{eq:H-def}) still appear to be in disorder, 
one can see by but straightforward calculations 
(see Appendix B) that the sum of these terms 
boils down to a form similar to the foregoing 
expression of $\calH_j$: 
\beq
  \lefteqn{
   p - \frac{1}{2}x_j\xi_j
       \Bigl(\sum_{k=1}^N \zeta(t_j - \lambda_k) 
            - \sum_{k\not= j}\zeta(t_j - t_k)\Bigr) 
     + \frac{1}{2}\sum_{k\not= j}\zeta(t_j -t_k)x_k\xi_k} 
  \nonumber \\
  &=& \sum_{k=1}^N 
      \frac{Q(t_j)P(\lambda_k)\sigma(t_j-\lambda_k+q)}
           {P'(t_j)Q'(\lambda_k)\sigma(t_j-\lambda_k)}
      (\zeta(t_j-\lambda_k+q) - \zeta(t_j-\lambda_k))\mu_k. 
  \label{eq:extra-terms}
\eeq

Combining these results, we obtain the following 
expression of $H_j$ in terms of the spectral 
Darboux coordinates: 
\beq
    H_j &=& 
    \sum_{k=1}^N 
    \frac{Q(t_j)P(\lambda_k)\sigma(t_j-\lambda_k+q)}
         {P'(t_j)Q'(\lambda_k)\sigma(t_j-\lambda_k)\sigma(q)}
    \times 
    \nonumber \\
    && \mbox{} 
    \times \Bigl(\mu_k^2 
      + (\zeta(t_j-\lambda_k+q)-\zeta(t_j-\lambda_k))\mu_k 
      - \sum_{l=1}^N C_l\wp(\lambda_k-t_l)\Bigr). 
\eeq
Eliminating $q$ by (\ref{eq:q-constraint}), 
we eventually obtain a final expression of 
the Hamiltonians: 
\beq
    H_j &=& 
    \sum_{k=1}^N 
    \frac{Q(t_j)P(\lambda_k)
          \sigma(\sum_{l\not= k}\lambda_l - \sum_{l\not= j}t_l)}
         {P'(t_j)Q'(\lambda_k)\sigma(t_j-\lambda_k)
          \sigma(\sum_{l=1}^N\lambda_l - \sum_{l=1}^Nt_l)}
    \times 
    \nonumber \\
    && \mbox{} 
    \times \Bigl(\mu_k^2 
      + (\zeta(t_j-\lambda_k+q)-\zeta(t_j-\lambda_k))\mu_k 
      - \sum_{l=1}^N C_l\wp(\lambda_k-t_l)\Bigr). 
    \label{eq:H-final}
\eeq

In summary, we have proven the following. 

\begin{theorem}
The isomonodromic $\rmSU(2)$ Calogero-Gaudin system 
can be converted to the non-autonomous Hamiltonian system 
\beq
  \frac{\rd\lambda_k}{\rd t_j} = \frac{\rd H_j}{\rd\mu_k}, \quad 
  \frac{\rd\mu_k}{\rd t_j} = - \frac{\rd H_j}{\rd\lambda_j} 
\eeq
in the spectral Darboux coordinates $\lambda_j,\mu_j$.  
The Hamiltonians are given by (\ref{eq:H-final}). 
\end{theorem}

\section{Relation to second order scalar ODE}

\subsection{Deriving second order ODE}

The structure of the Hamiltonians $H_j$ is very similar 
to Okamoto's Hamiltonians for isomonodromic deformations 
of a scalar ODE on a torus \cite{bib:Ok-ell2}.  
This is not a coincidence, but can be explained 
in the same way as the case of the $2 \times 2$ 
Schlesinger system \cite{bib:Ok-gar}.  

A clue is the fact that any $2 \times 2$ matrix system 
\beqnn
    \frac{dY}{dz} = L(z)Y, \quad 
    Y = \left(\begin{array}{c}
        y_1 \\
        y_2
        \end{array}\right),  
\eeqnn
yields a second order scalar ODE of the form 
\beq
    \frac{d^2y_1}{dz^2} + p_1(z)\frac{dy_1}{dz} + p_2(z)y_1 = 0. 
\eeq
The coefficients $p_1(z)$ and $p_2(z)$ of the latter 
are determined by the matrix $L(z)$ as follows: 
\beq
    p_1(z) &=& - \Tr L(z) - (\log L_{12}(z))', \\
    p_2(z) &=& \det L(z) - L_{11}'(z) + L_{11}(z)(\log L_{12}(z))'. 
\eeq
In our case, $L(z)$ is trace-free, so that 
the foregoing formulae of $p_1(z)$ and $p_2(z)$ 
become slightly simpler:  Firstly, $p_1(z)$ can be written 
explicitly as 
\beq
    p_1(z) = - (\log L_{12}(z))' 
           = - \sum_{j=1}^N \zeta(z - \lambda_j) 
             + \sum_{j=1}^N \zeta(z - t_j), 
    \label{eq:p1}
\eeq
which implies that $p_1(z)$ is doubly periodic. 
Secondly, $p_2(z)$ is also doubly periodic 
(as the quasi-periodicity of the matrix elements 
of $L(z)$ implies), and given by the formula 
\beqnn
    p_2(z) = - \frac{1}{2}\Tr L(z)^2 - L_{11}'(z) 
             + L_{11}(z)(\log L_{12}(z))'. 
\eeqnn
One can see from this formula that $p_2(z)$ has 
simple poles at $z = \lambda_j$ and double poles 
at $z = t_j$.  Let us express $p_2(z)$ as 
\beq
    p_2(z) = \sum_{j=1}^N \alpha_j \wp(z - t_j) 
           + \sum_{j=1}^N \beta_j \zeta(z - t_j) 
           + \sum_{j=1}^N \gamma_j \zeta(z - \lambda_j) 
           + \delta 
    \label{eq:p2-formula}
\eeq
and determine the coefficients by Laurent expansion 
at the poles.  
\begin{enumerate}
\item The first coefficient $\alpha_j$ can be read off 
from the $(z - t_j)^{-2}$ term of the Laurent expansion 
of $- \Tr L(z)^2/2$: 
\beq
    \alpha_j = - C_j = - \Tr A_j^2/2. 
\eeq
\item The second coefficient $\beta_j$ is the residue 
of $p_2(z)$ at $z = t_j$.  The term $L_{11}'(z)$ 
does not contribute to the residue.  The residue of 
the other terms at $z = t_j$ can be expressed as 
\beqnn
    - \Res_{z=t_j}\frac{1}{2}\Tr L(z)^2 = - \calH_j
\eeqnn
and 
\beqnn
  \Res_{z=t_j}L_{11}(z)(\log L_{12}(z))'
  &=& - p - \frac{1}{2}\sum_{k\not= j}\zeta(t_j-t_k)x_k\xi_k 
  \\
  && \mbox{} 
      + \frac{1}{2}x_j\xi_j
         \Bigl(\sum_{k=1}^N\zeta(t_j - \lambda_k) 
         - \sum_{k\not= j}\zeta(t_j - t_k) \Bigr). 
\eeqnn
As (\ref{eq:H-def}) shows, the sum of these two 
quantities is equal to $-H_j$.  Therefore 
\beq
    \beta_j = \Res_{z=t_j}p_2(z) = - H_j. 
\eeq
\item The thrid coefficient $\gamma_j$  is the residue 
of $p_2(z)$ at $z = \lambda_j$.  Since 
\beqnn
    - \Res_{z=\lambda_j}\frac{1}{2}\Tr L(z)^2 = 0 
\eeqnn
and 
\beqnn
    \Res_{z=\lambda_j}L_{11}(z)(\log L_{12}(z))' 
    = L_{11}(\lambda_j) 
    = \mu_j, 
\eeqnn
$\gamma_j$ can be expressed as 
\beq
    \gamma_j = \Res_{z=\lambda_j}p_2(z) = \mu_j. 
\eeq
\end{enumerate}
Thus the Hamiltonians $H_j$ and the ``momenta'' $\mu_j$ 
can be identified with the residues of $p_2(z)$.  
Exactly the same relation can be seen in the case 
of Garnier's isomonodromic system on a sphere 
\cite{bib:Ok-gar}.

\subsection{Another form of second order ODE}

Strictly speaking, however, the second order ODE above 
differs from that of Okamoto \cite{bib:Ok-ell2} and 
Kawai \cite{bib:Ka}, who consider a linear ODE 
of the form 
\beq
    \frac{d^2y}{dz^2} + p(z)y = 0. 
\eeq
At least formally, this discrepancy can be removed by 
the ``gauge transformation'' 
\beq
    y_1 = \exp\Bigl(-\frac{1}{2}\int^z p_1(z)dz\Bigr)y. 
\eeq
The coefficient $p(z)$ is given by 
\beq
    p(z) = - \frac{1}{2}p_1'(z) - \frac{1}{4}p_1(z)^2 + p_2(z). 
    \label{eq:p-def}
\eeq
Note, however, that this is actually a delicate procedure, 
because the gauge transformation might spoil 
the isomonodromic property.  Fortunately, 
the present case is free from this problem: 
The gauge transformation takes the form 
\beq
    y_1 = \left(\frac{\prod_{j=1}^N\sigma(z-\lambda_j)}
           {\prod_{j=1}^N\sigma(z-t_j)}\right)^{1/2}y, 
\eeq
and since the factor in front of $y$ has constant 
monodromy, the isomonodromic property is preserved 
by the gauge transformation. 

The zero-th order term $p(z)$ of the transformed ODE 
is a doubly periodic meromorphic function with 
second order poles at $z = \lambda_j,t_j$.  
The residues of $p(z)$ at these poles can be 
readily determined: 
\beq
    \Res_{z=\lambda_j}p(z) 
    &=& \mu_j - 2\sum_{k=1}^N\zeta(\lambda_j - t_k) 
        + 2\sum_{k\not= j}\zeta(\lambda_j - \lambda_k), 
    \\
    \Res_{z=t_j}p(z) 
    &=& - H_j - 2\sum_{k=1}^N\zeta(t_j - \lambda_k) 
        + 2\sum_{k\not= j}\zeta(t_j - t_k). 
\eeq
It is rather these quantities that Okamoto \cite{bib:Ok-ell2} 
and Kawai \cite{bib:Ka} use as Hamiltonians and conjugate 
variables of $\lambda_j$'s.  We can indeed reformulate 
our Hamiltonian system in that way.  Namely, if we define 
\beq
    \nu_j = \Res_{z=\lambda_j}p(z), \quad
    K_j = - \Res_{z=t_j}p(z), 
\eeq
these quantities satisfy the equation 
\beq
    \sum_{j=1}^N \mu_jd\lambda_j - \sum_{j=1}^N H_jdt_j 
  = \sum_{j=1}^N \nu_jd\lambda_j - \sum_{j=1}^N K_jdt_j 
    + \mbox{exact form}.  
\eeq
This implies that $\lambda_j$'s and $\nu_j$'s are 
a new set of Darboux coordinates, and that the previous 
Hamiltonian system is now converted to the new 
Hamiltonian system 
\beq
  \frac{\rd\lambda_k}{\rd t_j} = \frac{\rd K_j}{\rd\nu_k}, \quad 
  \frac{\rd\nu_k}{\rd t_j} = - \frac{\rd K_j}{\rd\lambda_j}. 
\eeq

\subsection{Reconstructing $2 \times 2$ matrix system}

Let us now consider the inverse problem.  
Namely, given the isomonodromic deformations of 
the second oder scalar ODE above, the problem is 
to reconstruct a $2 \times 2$ matrix system.  
A similar problem is discussed by Okamoto 
\cite[Section 3]{bib:Ok-gar} in the case of 
isomonodromic deformations on a sphere.  
In our case, the presence of diagonal gauge 
transformations (\ref{eq:diagonal-gt}) allows 
us to fix the coefficient $\kappa$ as $\kappa = 1$.  
Therefore we have only to show how to reconstruct 
the dynamical variables $q,p,A_k$'s of the matrix system 
from $\lambda_j,\mu_j$ of the scalar ODE.  

One can indeed reconstruct the $L$-matrix $L(z)$ 
by an algebraic procedure as follows.  
The first step is to reconsider (\ref{eq:L12-factor}) 
and (\ref{eq:q-constraint}) as {\it definition} 
of $L_{12}(z)$ and $q$.  The coefficient $\kappa$ 
is chosen to be $\kappa = 1$ as remarked above.  
$A_j^{-}$'s are thus determined.   Secondly, 
let $L_{11}(z)$ be a function of the form 
\beq
  L_{11}(z) = p + \sum_{j=1}^N \phi(q,z - t_j)A_j^3 
\eeq
that satisfy the interpolation relations 
$L_{11}(\lambda_j) = \mu_j$ for $j = 1,\ldots,N$.  
As we have seen in Section V, these relations 
can be solved for $A_j^3$ and $p$ under 
the constraint $\sum_{j=1}^N A_j^3 = 0$.  
As (\ref{eq:A-by-xxi}) suggests, $A_j^{+}$'s are 
to be determined as 
\beq
  A_j^{+} = \frac{\theta_j^2/4 - (A_j^3)^2}{A_j^{-}}.  
\eeq
One can thus reconstruct $L(z)$ from the scalar ODE.

\section{Conclusion}

We have applied the method of spectral Darboux 
coordinates to Korotkin and Samtleben's 
isomonodromic system on a torus \cite{bib:Ko-Sa}.  
The isomonodromic system has been thus converted to 
a non-autonomous Hamiltonian system in the spectral 
Darboux coordinates.  Although the Hamiltonians 
turn out to be a considerably intricate function, 
the method we have used is a rather straightforward 
analogue of the usual method for isomonodromic 
deformations on a sphere.  

Our non-autonomous Hamiltonian system may be thought 
of as an elliptic analogue of Garnier's isomonodromic 
systems \cite{bib:Ga1,bib:Ga2,bib:Ok-gar}.   
Almost the same system has been derived by Okamoto 
from isomonodromic deformations of a second order 
scalar ODE on a torus \cite{bib:Ok-ell2}.  We have seen 
how these two systems are related.  Speaking differently, 
our approach from a $2 \times 2$ matrix system reveals 
a hidden algebro-geometric meaning of the Hamiltonian 
structure in Okamoto's work \cite{bib:Ok-gar}. 

An important lesson of the present work is that 
the notions of spectral curve and spectral 
Darboux coordinates persist to be useful and 
essential beyond isospectral deformations.  
This observation lies in the heart of the work 
of Harnad and Wisse \cite{bib:Ha-Wi}.  We have 
confirmed it for an example of isomonodromic 
deformations on a torus.  

In this respect, an interesting problem is to 
describe  the isomonodromic $\rmSU(2)$ pure Gaudin 
system \cite{bib:Ta-gau,bib:Ko-Ma-Sa} from the same 
point of view.  Separation of variables of 
the isospectral partner has been studied by 
Sklyanin and Takebe \cite{bib:Sk-Ta} (see also 
the paper of Hurtubise and Kjiri \cite{bib:Hu-Kj} 
for geometric aspects).  The work of Sklyanin and Takebe 
shows that separation of variables of this system is 
technically far more complicated than the Calogero-Gaudin 
system.  This will be also the case for the isomonodromic 
analogue.  

Let us conclude the present consideration with 
a remark on trigonometric and rational analogues. 
The trigonometric and rational analogues of 
Korotkin and Samtleben's isomonodromic deformations 
can be obtained by replacing the basic functions $\sigma(z)$, 
$\zeta(z)$ and $\phi(u,z)$ by the following trigonometric 
or rational functions: 
\begin{enumerate}
\item Trigonometric model 
\beq
    \sigma(z) = \sin z, \quad 
    \zeta(z) = \frac{\cos z}{\sin z}, \quad 
    \phi(u,z) = \frac{\cos z}{\sin z} - \frac{\cos u}{\sin u}. 
\eeq
\item Rational model
\beq
    \sigma(z) = z, \quad 
    \zeta(z) = \frac{1}{z}, \quad 
    \phi(u,z) = \frac{1}{z} - \frac{1}{u}.  
\eeq
\end{enumerate}
A hyperbolic model will be obtained if one replaces the 
trigonometric functions by the corresponding hyperbolic 
functions.  These are nothing but the well known pattern 
of degeneration of the Calogero-Moser systems; 
the Calogero-Gaudin systems, too, obey this pattern.  
In fact, it is the rational model in this list that 
Brzezi\'nski considered in his work \cite{bib:Br}.  
One can formulate an isomonodromic partner of 
these degenerate Calogero-Gaudin systems as in 
the case of the elliptic model.  Presumably, 
those isomonodromic systems will not be known 
in the literature.

\subsection*{Acknowledgements}

I am grateful to Shingo Kawai and Takashi Takebe for 
many useful comments.  This work is partially supported by 
the Grant-in-Aid for Scientific Research (No. 12640169) 
from the Ministry of Education, Culture, Sports and 
Technology.

\startappendix

\section{Interpolation formula}

Let us examine the auxiliary function 
\beqnn
    f_k(z) = \frac{Q(z)\sigma(z - \lambda_k + q)}
                  {P(z)\sigma(z - \lambda_k)}.  
\eeqnn
This is a doubly periodic meromorphic function 
with simple zeros at $\lambda_j$ ($j \not= k$) and 
$\lambda_k - q$ and simple poles at $t_j$ 
($j = 1,\ldots,N$).  By the residue theorem, 
the residues 
\beqnn
    \Res_{z=t_j} f_k(z) 
    = \frac{Q(t_j)\sigma(t_j - \lambda_k + q)}
           {P'(t_j)\sigma(t_j - \lambda_k)} 
\eeqnn
at the poles $z = t_j$ obey the sum-to-zero 
constraint 
\beq
     \sum_{j=1}^N 
     \frac{Q(t_j)\sigma(t_j - \lambda_k + q)}
          {P'(t_j)\sigma(t_j - \lambda_k)} 
     = 0. 
     \label{eq:sum-to-zero}
\eeq
Let us consider the linear combination 
\beqnn
      \sum_{j=1}^N
      \frac{Q(t_j)\sigma(t_j - \lambda_k + q)}
      {P'(t_j)\sigma(t_j - \lambda_k)} \zeta(z - t_j) 
\eeqnn
of $\zeta(z - t_j)$ weighted by these residues. 
Since this function is a doubly periodic meromorphic 
function with the same set of simple poles and residues as 
$f_k(z)$, it differs from $f_k(z)$ by at most a constant: 
\beqnn
    \frac{Q(z)\sigma(z - \lambda_k + q)}
         {P(z)\sigma(z - \lambda_k)}
    = \sum_{j=1}^N
      \frac{Q(t_j)\sigma(t_j - \lambda_k + q)}
      {P'(t_j)\sigma(t_j - \lambda_k)} \zeta(z - t_j) 
    + \mbox{constant}. 
\eeqnn
Moreover, since the left hand side vanishes 
at $z = \lambda_k - q$, the constant term on 
the right hand side can be easily determined 
as follows: 
\beqnn
    \mbox{constant} 
    = - \sum_{j=1}^N 
        \frac{Q(t_j)\sigma(t_j - \lambda_k + q)}
        {P'(t_j)\sigma(t_j - \lambda_k)}
        \zeta(\lambda_k - t_j - q). 
\eeqnn
We thus obtain the interpolation formula 
\beq
    \frac{Q(z)\sigma(z - \lambda_k + q)}
         {P(z)\sigma(z - \lambda_k)}
    = \sum_{j=1}^N
      \frac{Q(t_j)\sigma(t_j - \lambda_k + q)}
           {P'(t_j)\sigma(t_j - \lambda_k)} 
      (\zeta(z - t_j) - \zeta(\lambda_k - t_j - q)). 
    \label{eq:interpol}
\eeq

One can derive the following three identities 
from this interpolation formula. 

\begin{enumerate}
\item Since the left hand side of the interpolation formula 
vanishes at $z = \lambda_l$ ($l \not= k$), 
\beq
    \sum_{j=1}^N
    \frac{Q(t_j)\sigma(t_j - \lambda_k + q)}
         {P'(t_j)\sigma(t_j - \lambda_k)} 
    (\zeta(\lambda_l - t_j) - \zeta(\lambda_k - t_j - q))
    = 0   \quad (l \not= k).  
    \label{eq:interpol-1}
\eeq
\item  By letting $z \to \lambda_k$ in the interpolation 
formula, 
\beq
    \sum_{j=1}^N 
    \frac{Q(t_j)\sigma(t_j - \lambda_k + q)}
         {P'(t_j)\sigma(t_j - \lambda_k)} 
    (\zeta(\lambda_k - t_j) - \zeta(\lambda_k - t_j - q))
    = \frac{Q'(\lambda_k)\sigma(q)}{P'(\lambda_k)}. 
    \label{eq:interpol-2}
\eeq
\item  By replacing $k \to l$, $j \to k$ and 
separating a term from the sum, the interpolation 
formula takes the form 
\beqnn
  \lefteqn{\sum_{k\not= j}
     \frac{Q(t_k)\sigma(t_k - \lambda_l + q)}
          {P'(t_k)\sigma(t_k - \lambda_l)}\zeta(z - t_k)} 
  \\
  &=& \frac{Q(z)\sigma(z - \lambda_l + q)}
           {P(z)\sigma(z - \lambda_l)} 
    - \frac{Q(t_j)\sigma(t_j - \lambda_l + q)}
           {P'(t_j)\sigma(t_j - \lambda_l)}\zeta(z - t_j) 
  \\
  && \mbox{} 
     + \sum_{k=1}^N 
       \frac{Q(t_k)\sigma(t_k-\lambda_l+q)}
            {P'(t_k)\sigma(t_k-\lambda_l)}
       \zeta(\lambda_l-t_k-q). 
\eeqnn
By letting $z \to t_j$, 
\beq
  \lefteqn{\sum_{k\not= j}
     \frac{Q(t_k)\sigma(t_k - \lambda_l + q)}
          {P'(t_k)\sigma(t_k - \lambda_l)}\zeta(t_j - t_k)}
  \nonumber \\
  &=& \frac{Q(t_j)\sigma(t_j - \lambda_l + q)}
           {P'(t_j)\sigma(t_j - \lambda_\ell)}
      \Bigl(- \frac{1}{2}\frac{P''(t_j)}{P'(t_j)} 
            + \frac{Q'(t_j)}{Q(t_j)} 
            - \zeta(t_j - \lambda_l)
            + \zeta(t_j - \lambda_l + q) \Bigr) 
  \nonumber\\
  && \mbox{} 
     + \sum_{k=1}^N 
       \frac{Q(t_k)\sigma(t_k - \lambda_l + q)}
            {P'(t_k)\sigma(t_k - \lambda_l)}
       \zeta(\lambda_l - t_k - q). 
     \label{eq:interpol-3}
\eeq
\end{enumerate}

\section{Calculation of extra terms in (\ref{eq:H-def})}

Let us use (\ref{eq:xxi-by-lamu}) to rewrite the 
last piece on the right hand side of (\ref{eq:H-def}) as 
\beqnn
    \frac{1}{2}\sum_{k=1}^N \zeta(t_j-t_k)x_k\xi_k 
    = \sum_{l=1}^N\Bigl( 
        \sum_{k\not= j}
        \frac{Q(t_k)\sigma(t_k-\lambda_l+q)}
             {P'(t_k)\sigma(t_k-\lambda_l)}\zeta(t_j-t_k) \Bigr)
      \frac{P(\lambda_l)\mu_l}{Q'(\lambda_l)\sigma(q)}. 
\eeqnn
The sum over $k\not= j$ arising here has been 
partially calculated in (\ref{eq:interpol-3}).  
Using the identities 
\beqnn
    \frac{1}{2}\frac{P''(t_j)}{P'(t_j)} 
      = \sum_{k\not= j} \zeta(t_j-t_k), \quad 
    \frac{Q'(t_j)}{Q(t_j)} 
      = \sum_{k=1}^N \zeta(t_j-\lambda_k) 
\eeqnn
on the right hand side of (\ref{eq:interpol-3}), 
one can rewrite the foregoing quantity as 
\beqnn
  \lefteqn{\frac{1}{2}\sum_{k=1}^N \zeta(t_j-t_k)x_k\xi_k} \\
  &=& 
      \sum_{l=1}^N 
      \frac{Q(t_j)P(\lambda_l)\sigma(t_k-\lambda_l+q)\mu_\ell}
           {P'(t_j)Q'(\lambda_l)\sigma(t_j-\lambda_l)\sigma(q)}
      \Bigl(- \sum_{k\not= j}\zeta(t_j-t_k) 
            + \sum_{k=1}^N \zeta(t_j-\lambda_k)\Bigr) \\
  && \mbox{}
    + \sum_{l=1}^N 
      \frac{Q(t_j)P(\lambda_l)\sigma(t_k-\lambda_l+q)}
           {P'(t_j)Q'(\lambda_l)\sigma(t_j-\lambda_l)\sigma(q)}
      (\zeta(t_j-\lambda_l+q) - \zeta(t_j-\lambda_l))\mu_\ell \\
  && \mbox{}
    + \sum_{k,l=1}^N 
        \frac{Q(t_k)P(\lambda_l)\sigma(t_k-\lambda_l+q) 
              \zeta(\lambda_l-t_k-q)\mu_l}
             {P'(t_k)Q'(\lambda_l)\sigma(t_k-\lambda_l)\sigma(q)}. 
\eeqnn
By (\ref{eq:xxi-by-lamu}) and (\ref{eq:p-by-lamu}), 
the first and third lines on the right hand side turn 
into the following form: 
\beqnn
   \mbox{first line} &=& 
   - \frac{1}{2}x_j\xi_j\Bigl(
         \sum_{k\not= j}\zeta(t_j-t_k) 
          - \sum_{k=1}^N\zeta(t_j-\lambda_k)\Bigr), \\
   \mbox{third line} &=& -p. 
\eeqnn
Since the sum of these two cancels the second and 
third pieces on the right hand side of (\ref{eq:H-def}), 
we eventually obtain the identity 
\beqnn
  \lefteqn{
   p - \frac{1}{2}x_j\xi_j
       \Bigl(\sum_{k=1}^N \zeta(t_j - \lambda_k) 
            - \sum_{k\not= j}\zeta(t_j - t_k)\Bigr) 
     + \frac{1}{2}\sum_{k\not= j}\zeta(t_j -t_k)x_k\xi_k} 
  \\ 
  &=& \sum_{l=1}^N 
      \frac{Q(t_j)P(\lambda_l)\sigma(t_j-\lambda_l+q)}
           {P'(t_j)Q'(\lambda_l)\sigma(t_j-\lambda_l)\sigma(q)}  
      (\zeta(t_j-\lambda_l+q) - \zeta(t_j-\lambda_l))\mu_l, 
\eeqnn
which is nothing but (\ref{eq:extra-terms}).

\section{Zero-curvature equations in more detail}

Since the diagonal part of the zero-curvature equation 
(\ref{eq:zc}) has been specified in Section III, 
let us now examine the off-diagonal part.  

As regards the upper right part, the matrix element of 
the commutator $[\widetilde{M}_j(z),\widetilde{M}_k(z)]$ reads 
\beqnn
  [\widetilde{M}_j(z),\widetilde{M}_k(z)]_{12} 
  &=& 2(p_j + \zeta(z-t_j)A_j^3)\phi(q,z-t_k)A_k^{-} \\
  &&\mbox{} 
    - 2\phi(q,z-t_j)A_j^{-}(p_k + \zeta(z-t_k)A_k^3). 
\eeqnn
One can use the functional identity 
(\ref{eq:phi(u,z)phi(-u,z)}) in the form 
\beqnn
  \phi(u,z-w)\zeta(z) 
  = - \phi(u,z)\phi(u,-w) + \phi(u,z-w)\zeta(w) 
    - \phi_u(u,z-w) 
\eeqnn
to eliminate $\zeta(z-t_j)\phi(q,z-t_k)$ and 
$\phi(q,z-t_j)\zeta(z-t_k)$ as 
\beqnn
  [\widetilde{M}_j(z),\widetilde{M}_k(z)]_{12} 
  &=& 2\phi(q,z-t_k)(p_jA_k^{-} + \zeta(t_k-t_j)A_j^3A_k^{-} 
        + \phi(q,t_k-t_j)A_j^{-}A_k^3) \\
  &&\mbox{}
    - 2\phi(q,z-t_j)(p_kA_j^{-} + \zeta(t_j-t_k)A_j^{-}A_k^3 
        + \phi(q,t_j-t_k)A_j^3A_k^{-}) \\
  &&\mbox{} 
    - 2\phi_u(q,z-t_k)A_j^3A_k^{-} 
    + 2\phi_u(q,z-t_j)A_j^{-}A_k^3. 
\eeqnn	
On the other hand, 
the derivative part of the zero-curvature equation 
can be expressed as 
\beqnn
\lefteqn{
  \frac{\rd\widetilde{M}_{k,12}(z)}{\rd t_j} 
  - \frac{\rd\widetilde{M}_{j,12}(z)}{\rd t_k}
}\\
  &=& \phi(q,z-t_k)\frac{\rd A_k^{-}}{\rd t_j} 
    + \phi_u(q,z-t_k)\frac{\rd q}{\rd t_j}A_k^{-} 
    - \phi(q,z-t_j)\frac{\rd A_j^{-}}{\rd t_k} 
    - \phi_u(q,z-t_j)\frac{\rd q}{\rd t_k}A_j^{-}. 
\eeqnn
Thus the $\phi_u$ terms cancel out in the zero-curvature equation 
and one is left with an equation of the form 
\beqnn
  \phi(q,z-t_j)(\cdots) + \phi(q,z-t_k)(\cdots) = 0 
\eeqnn
with the coefficients $(\cdots)$ that do not depend on $z$.  
As it turns out, these coefficients are exactly the same 
as some of the equations of motion of $q$, $A_j$ and $A_k$.  
One can thus see that this part of the zero-curvature 
equations is automatically satisfied.  

The lower left part of the zero-curvature equation can be 
treated in the same way.

\section{Integrability of (\ref{eq:zc-pj})}

In components, the closedness condition $d\omega = 0$ reads 
\beq
  \rd_{t_j}\Bigl(
    \phi(q,t_j-t_k)(\zeta(t_j-t_k-q) + \zeta(q))A_j^{+}A_k^{-}
  \Bigr) \nonumber \\ 
  + (\mbox{cyclic permutations of $j,k,\ell$}) = 0. 
  \label{eq:closedness}
\eeq
The goal is to show that these equations are indeed 
satisfied under the equations of motion (\ref{eq:eom-qp}), 
(\ref{eq:eom-Apm}), (\ref{eq:eom-A3}) of $q,p$ and $A$'s.  
Applying the Leibniz rule to the left hand side of 
(\ref{eq:closedness}) yields such terms as 
\beqnn
  - \phi(q,t_j-t_k)\Bigl(
      (\zeta(t_j-t_k-q) + \zeta(q))^2 
      + \zeta'(t_j-t_k-q) + \zeta'(q) 
    \Bigr)\frac{\rd q}{\rd t_j}A_j^{+}A_k^{-} \\
  + \phi(q,t_j-t_k)(\zeta(t_j-t_k-q) + \zeta(q)) 
    \Bigl(\frac{\rd A_j^{+}}{\rd t_j}A_k^{-} 
      + A_j^{+}\frac{\rd A_k^{-}}{\rd t_j}\Bigr) 
\eeqnn
and their cyclic permutations.  One can eliminate 
the derivatives of $q,p$ and $A$'s by the equations 
of motion; this in turn yields linear and quadratic 
combinations of $\phi$'s.  As regards the quadratic 
combinations, one can use the functional identity 
(\ref{eq:phi(u,z)phi(-u,z)}) to reduce such terms 
to a linear combinations of $\phi$'s, e.g., 
\beqnn
  \phi(q,t_j-t_k)\phi(q,t_\ell-t_j) 
  = \phi(q,t_\ell-t_k)\Bigl(
        \zeta(q-t_\ell+t_k) - \zeta(q) 
        + \zeta(t_j-t_k) + \zeta(t_\ell-t_j)\Bigr), 
\eeqnn
etc.  After some more algebra, one can thus eventually 
convert (\ref{eq:closedness}) to equations of the form 
\beq
  \phi(q,t_j-t_k)c_{jk\ell}A_j^{+}A_k^{-}A_\ell^3 
  + (\mbox{cyclic permutations of $j,k,\ell$}) 
  = 0 
\eeq
where 
\beqnn
  c_{jk\ell} 
  &=& (\zeta(t_j-t_k-q) + \zeta(q))^2 
      + \zeta'(t_j-t_k-q) + \zeta'(q) \nonumber \\
  &&\mbox{} 
      + (\zeta(t_j-t_k-q) + \zeta(q))
        (\zeta(t_\ell-t_j) - \zeta(t_\ell-t_k))  \nonumber \\
  &&\mbox{} 
      + (\zeta(t_\ell-t_k-q) + \zeta(q)) 
        (- \zeta(t_j-t_k-q) - \zeta(q) 
         + \zeta(t_\ell-t_k) + \zeta(t_j-t_\ell)) \nonumber \\
  &&\mbox{} 
      + (\zeta(t_j-t_\ell-q) + \zeta(q)) 
        (- \zeta(t_j-t_k-q) - \zeta(q) 
         + \zeta(t_j-t_\ell) + \zeta(t_\ell-t_k)). 
\eeqnn
Actually, these coefficients $c_{jk\ell}$ turn out 
to vanish identically.  One can indeed verify that 
the function 
\beqnn
  f_{jk\ell}(z)
  &=& (\zeta(t_j-t_k-z) + \zeta(z))^2 
      + \zeta'(t_j-t_k-z) + \zeta'(z) \nonumber \\
  &&\mbox{} 
      + (\zeta(t_j-t_k-z) + \zeta(z))
        (\zeta(t_\ell-t_j) - \zeta(t_\ell-t_k))  \nonumber \\
  &&\mbox{} 
      + (\zeta(t_\ell-t_k-z) + \zeta(z)) 
        (- \zeta(t_j-t_k-z) - \zeta(z) 
         + \zeta(t_\ell-t_k) + \zeta(t_j-t_\ell)) \nonumber \\
  &&\mbox{} 
      + (\zeta(t_j-t_\ell-z) + \zeta(z)) 
        (- \zeta(t_j-t_k-z) - \zeta(z) 
         + \zeta(t_j-t_\ell) + \zeta(t_\ell-t_k)) 
\eeqnn
of the complex variable $z$ is a doubly periodic entire 
function with a zero at $z = (t_j-t_k)/2$; 
this implies that $f_{jk\ell}(z) = 0$.

\end{document}